\documentclass[12pt]{article}
\usepackage[utf8]{inputenc}
\usepackage{amssymb, amsmath}
\usepackage[dvipdfmx]{graphicx}
\usepackage{xcolor}
\usepackage[colorlinks=true ,urlcolor=blue ,anchorcolor=blue ,citecolor=blue ,filecolor=blue ,linkcolor=blue ,menucolor=blue ,linktocpage=true ,pdfproducer=medialab ,pdfa=true]{hyperref}

\begin{document}

\thispagestyle{empty}
\begin{flushright}
April 2023 \\
RESCEU-9/23
\end{flushright}
\vspace*{1cm}
\begin{center}

{\Large \bf Baryogenesis from sphaleron decoupling}

\vspace*{.55cm} {\large Muzi Hong$^{a,b }$\footnote{\tt
hong@resceu.s.u-tokyo.ac.jp}, Kohei Kamada$^{b }$\footnote{\tt
kohei.kamada@resceu.s.u-tokyo.ac.jp} and Jun'ichi Yokoyama$^{a,b,c,d}$\footnote{\tt
yokoyama@resceu.s.u-tokyo.ac.jp}
}\\
\vspace{.5cm} {\em $^a$Department of Physics, Graduate School of Science, The University of Tokyo}\\
\vspace{.05cm}{\em Hongo 7-3-1 Bunkyo-ku, Tokyo 113-0033, Japan}\\
\vspace{.3cm}
{\em $^b$Research Center for the Early Universe (RESCEU), Graduate School of Science}\\
\vspace{.05cm}
{\em The University of Tokyo, Hongo 7-3-1 Bunkyo-ku, Tokyo 113-0033, Japan}\\
\vspace{.3cm}
{\em $^c$Kavli Institute for the Physics and Mathematics of the Universe (Kavli IPMU), UTIAS, WPI}\\
\vspace{.05cm}
{\em The University of Tokyo, Kashiwa, Chiba, 277-8568, Japan}\\
\vspace{.3cm}
{\em $^d$Trans-scale Quantum Science Institute}\\
\vspace{.05cm}
{\em The University of Tokyo, Hongo 7-3-1 Bunkyo-ku, Tokyo 113-0033, Japan
}\\

\vspace{1cm} ABSTRACT
\end{center}
\hspace{\parindent}
The electroweak sphaleron process 
breaks the baryon number conservation within the realms of the Standard Model of particle physics (SM). 
Recently, it is pointed out that its decoupling may provide the out-of-equilibrium condition required for baryogenesis. 
In this paper, we study such a scenario 
taking into account the baryon-number wash-out effect of the sphaleron itself to improve the estimate. 
We clarify the amount of CP violation required for this scenario to explain the observed asymmetry.

\vfill \setcounter{page}{0} \setcounter{footnote}{0}

\vspace{1cm}
\newpage

\section{Introduction}

Baryogenesis, the origin of baryon asymmetry of the Universe (BAU), is yet an unsolved problem. Sakharov \cite{Sakharov:1967dj} identified 
three necessary conditions for baryogenesis: (1) baryon-number non-conservation, (2) C- and CP-violation, and (3) deviation from thermal equilibrium. Various scenarios (e.g. \cite{Kuzmin:1985mm,Fukugita:1986hr,Affleck:1984fy};  for recent review, see Ref.~\cite{Bodeker:2020ghk}) satisfying these conditions have been proposed to explain the observed baryon-to-entropy
ratio \cite{Planck:2018nkj,ParticleDataGroup:2022pth}

\begin{equation}
\label{observationnb}
\frac{n_B}{s}\simeq 9\times 10^{-11}, 
\end{equation} 
with $n_B$ and $s$ being the baryon number density and entropy density, respectively.
All scenarios that reproduce this value known to date are based on physics beyond the Standard Model of particle physics (BSM). Electroweak (EW) baryogenesis \cite{Kuzmin:1985mm} and leptogenesis \cite{Fukugita:1986hr} are some of the promising examples. EW baryogenesis requires a more complicated Higgs sector, so that CP violation can be enhanced and a first order EW phase transition (EWPT) instead of a smooth cross over~\cite{DOnofrio:2015gop}, as in the Standard Model of particle physics (SM) \cite{Buchmuller:1994qy,Kajantie:1996mn,Csikor:1998eu}
with the Higgs boson mass around 125 GeV~\cite{ATLAS:2012yve,CMS:2012qbp}, can appear to create an inequilibrium environment, while leptogenesis requires right-handed Majorana neutrinos whose decay~\cite{Fukugita:1986hr} or oscillation~\cite{Akhmedov:1998qx,Asaka:2005pn} provides
the inequilibrium process.

Sphaleron process \cite{Manton:1983nd,Klinkhamer:1984di,Arnold:1987mh} plays a key role in both electroweak baryogenesis and leptogenesis, because it breaks baryon-number conservation. It changes the gauge field configuration topologically, resulting in the violation of baryon $B$ and lepton $L$ charges~\cite{tHooft:1976rip} due to the chiral anomaly~\cite{Adler:1969gk,Bell:1969ts}, while conserving $B-L$. Specifically, an energy barrier exists between topologically different vacua in the space of field configuration \cite{Jackiw:1976pf}, and sphaleron process describes the barrier crossing between different vacua through thermal fluctuations. Each topologically distinct vacuum is characterized by an integer $N_\mathrm{CS}$, and the sphaleron process increasing it by unity will generate 9 quarks (3 baryons), while the one decreasing it by unity will generate 9 anti-quarks ($-3$ baryons)~\cite{tHooft:1976rip}. 

Sphaleron process can occur within the SM. However, it has been believed that CP violation in the SM is insufficient to generate enough baryon number \cite{Shaposhnikov:1987tw,Farrar:1993sp,Farrar:1993hn,Gavela:1993ts,Gavela:1994dt}, and the smooth cross over at EW symmetry breaking~\cite{Buchmuller:1994qy,Kajantie:1996mn,Csikor:1998eu} in the SM with the 125 GeV Higgs~\cite{ATLAS:2012yve,CMS:2012qbp} does not realize the required inequilibrium environment, and hence the EW baryogenesis
does not work in the SM. 
It would be appealing to explain baryon asymmetry within the SM instead of using new physics, and it is important to see if it is truly impossible to solve this problem using known physics.

Recently, Kharzeev et al. \cite{Kharzeev:2019rsy} proposed an interesting scenario, trying to realize baryogenesis within the SM. They argued that when Higgs boson acquires a nonvanishing vacuum expectation value (VEV), 
the energy barrier acquires a scale, and  
various size of sphaleron-like configurations 
with different energy contribute to the baryon-number violating
transition. 
Sphalerons
start decoupling in sequence according to their energy, providing out-of-equilibrium condition required for baryogenesis, and this sphaleron decoupling process continues until sphaleron processes  freeze out completely. 
The number of quarks and that of anti-quarks generated during those decoupling process are different due to the CP violation originated from the Cabbibo-Kobayashi-Maskawa (CKM) matrix, and they claimed that, unlike the common belief~\cite{Shaposhnikov:1987tw,Farrar:1993sp,Farrar:1993hn,Gavela:1993ts,Gavela:1994dt},  
it is possible to pick up sufficient CP violation from CKM matrix and this provides a source of baryon number. Their estimate showed the baryon-to-photon ratio generated from this source is only one order of magnitude smaller than the observed value. 

In this article, we study the scenario more quantitatively
by formulating a kinetic equation for the BAU.  We point out that their estimation did not involve the wash-out effect of baryon number due to the sphalerons that are still in equilibrium. 
We give the criteria of which sizes of sphaleron-like 
configurations are in or out-of equilibrium to incorporate
wash-out effect in the kinetic equation.
We determine the required amount of CP violation for this scenario to explain the observed BAU, 
which turns out to be two to three  orders of magnitude larger than that evaluated in Ref.~\cite{Kharzeev:2019rsy} for the SM.

The paper is organized as follows. In the next section,  
we give a brief review of the EW sphaleron and baryogenesis 
from its decoupling studied
in Ref.~\cite{Kharzeev:2019rsy}.  We then formulate
the kinetic equation for the baryon number 
by identifying 
the source term and the wash-out term around the sphaleron decoupling temperature. In Sec.~\ref{sec:result},   
we solve it numerically to give 
an estimate for the resulting BAU. 
We give our concluding remarks in Sec.~\ref{sec:conclusion}. 

\section{Formulation of the kinetic equation} 

\subsection{Baryon number violation in the SM}
Let us start with reviewing 
the baryon-number violation in the SM. 
The baryon number in a system changes through the chiral anomaly~\cite{Adler:1969gk,Bell:1969ts}
when the quantum tunneling process between the topologically different 
degenerate vacua takes place.
At zero temperature, 
weak SU(2) instanton which mediates such a tunneling
is found to be exponentially suppressed, $[\exp(- 8 \pi^2/g^2)]^2 \sim 10^{-173}$~\cite{tHooft:1976rip}. 
While this guarantees the stability of the BAU,  
this process would not be its origin. 

At finite temperature, baryon-number violation 
takes place by a transition over the energy barrier between topologically different vacua through thermal fluctuations. When the temperature becomes lower than the EW scale, the energy barrier gains a scale
determined by the Higgs expectation value.
The transition proceeds by crossing around
the field configuration at the top of the energy barrier, which is called a sphaleron, 
and this transition process is called a sphaleron process.

The sphaleron is a spherically symmetric static  
solution of the field equation of the SU(2)-Higgs theory\footnote{In the SM, hyper U(1) gauge field is also involved through the non-zero weak mixing angle, but its effect has been turned out to give just a perturbative modification~\cite{Klinkhamer:1984di}.}.
The ansatz for the sphaleron configuration adopted in Ref.~\cite{Klinkhamer:1984di} is given as follows. 
Let us introduce profile functions, 
$f(\xi)$ and $h(\xi)$, which describe the relevant physical degrees of freedom 
of the SU(2) gauge field $W_i^a$ and the Higgs doublet field $\varphi$ as 
\begin{equation}
    W_i^a \sigma^a \mathrm{d}x^i = -\frac{2\mathrm{i}}{g} f(\xi) \mathrm{d}U^\infty (U^\infty)^{-1}, \quad \varphi = \frac{v}{\sqrt{2}} h(\xi) U^\infty \left(\begin{array}{c} 0 \\1 \end{array}\right), \quad i=1,2,3, 
\end{equation}
with $\sigma^a$, $g$ , and $v$ being the Pauli matrices, the SU(2) coupling constant, and the vacuum expectation value (VEV) of
the Higgs field, respectively. $U^\infty$ denotes a two dimensional matrix
\begin{equation}
    U^\infty = \frac{1}{r} \left(\begin{array}{cc} x^3 & x^1 + \mathrm{i} x^2 \\ - x^1+ \mathrm{i} x^2 & x^3 \end{array}\right), \quad r\equiv \sqrt{\sum_i {x^i}^2}, \quad \xi \equiv g v r. 
\end{equation}
Here we have adopted the gauge fixing $W_0=0$. Reference~\cite{Klinkhamer:1984di} 
found that the following ansatz, 
\begin{equation} \label{sphaleronconff}
f(\xi)=
\begin{cases}
\dfrac{\xi^2}{\Xi(\Xi+4)} & \xi\leq\Xi \\
1-\dfrac{4}{\Xi+4}\textrm{exp}\left[\dfrac{1}{2}(\Xi-\xi)\right] & \xi\geq\Xi,
\end{cases}
\end{equation}
\begin{equation} \label{sphaleronconfh}
h(\xi)=
\begin{cases}
\dfrac{\sigma\Omega+1}{\sigma\Omega+2}\dfrac{\xi}{\Omega} & \xi\leq\Omega \\
1-\dfrac{\Omega}{\sigma\Omega+2}\dfrac{1}{\xi}\textrm{exp}\left[\sigma(\Omega-\xi)\right] & \xi\geq\Omega,
\end{cases}
\end{equation}
fit the numerical solutions relatively well, which we will use in the following.
Here $\sigma=\sqrt{2\lambda/g^2}$ ($\lambda$ is the Higgs quartic coupling) and the boundary conditions
\begin{equation}
    f(0) = h(0) = 0, \quad \lim_{r \rightarrow \infty} f(r) = \lim_{r \rightarrow \infty} h(r) = 1, 
\end{equation}
are imposed.
The values of $\Xi$ and $\Omega$ are determined by the minimization condition 
of the sphaleron mass (or energy)~\cite{Kajantie:1995dw}, 
\begin{equation} \label{sphaleronmass}
    M=\frac{4\pi v}{g}\int_0^\infty \mathrm{d}\xi\left[4(f^\prime)^2+\frac{8}{\xi^2}f^2(1-f)^2+\frac{\xi^2}{2}(h^\prime)^2+h^2(1-f)^2+\frac{\xi^2}{16} \sigma^2
    (h^2-1)^2\right], 
\end{equation}
where the prime denotes the derivative with respect to $\xi$. 
At finite temperature, we use the 
coupling constants as well as the 
Higgs expectation value evaluated at the temperature scale of interest.
Around the EW scale, for Higgs boson mass around 125 GeV
we extrapolate the results for the three-dimensional effective theory~\cite{Kajantie:1995dw} as
\begin{equation}
    \frac{{\bar \lambda}_3}{{\bar g}_3^2}\approx0.22,\ \ \ \ {\bar g}_3^2\approx0.39,
    \label{lambda}
\end{equation} 
and find numerically $\Xi_0 = 1.467$ and $\Omega_0=1.701$.
Later we also consider sphaleron-like configurations parameterized by a single parameter $a$ with $\Xi = a \Xi_0$ and $\Omega= a \Omega_0$. Since $\Xi$ and $\Omega$ represent the size of SU(2) gauge field and Higgs field, respectively, $a$ means the size of sphaleron-like configurations.

Analytically, to evaluate the transition rate per unit time and unit volume, or the sphaleron rate, below EW temperature, one calculates the ensemble average of ``probability current'' at the top of the energy barrier, by taking path-integral around the sphaleron background. The sphaleron rate 
is given as~\cite{Arnold:1987mh}
\begin{equation} \label{sphaleronrateanal1}
 \Gamma_\mathrm{sph} \sim T W \mathrm{Det'} \{3\} \exp[-M(T)/T], 
\end{equation}
where $\mathrm{Det'} \{3\} $ is a three dimensional determinant
obtained by integrating fluctuations
around the 
sphaleron background and $W$ is the volume factor coming from the 
zero-modes associated with the translational invariance. 
$M(T)$ is the temperature-dependent sphaleron mass 
calculated using Eq.~\eqref{sphaleronmass}, 
evaluated with the 
temperature-dependent Higgs VEV, $v(T)$. 
At temperatures higher than the EW scale, 
the energy barrier for the sphaleron no longer exists, 
but the baryon-number violating transition 
process is governed by the 
parameter $g^2 T$. 
From the dimensional estimate, 
the sphaleron rate is evaluated as~\cite{Arnold:1987mh,Arnold:1996dy}
\begin{equation}\label{sphaleronrateanal2}
    \Gamma_\mathrm{sph} \sim \alpha_\mathrm{W}^5 T^4, 
\end{equation}
where $\alpha_\mathrm{W} \equiv g^2/4\pi$.
Here the additional factor $\alpha_\mathrm{W}$ is due to the plasma 
damping effect~\cite{Arnold:1996dy}.

Since it is difficult to go beyond the approximate formulas (\ref{sphaleronrateanal1}) and (\ref{sphaleronrateanal2}) calculating the precise values of their coefficients analytically, numerical analysis is often performed but from a different perspective, namely, based on the diffusion equation of the Chern-Simons number, which yields 
\begin{equation} \label{diffrate}
\Gamma_\mathrm{sph} (T) \equiv \lim_{V,t \rightarrow \infty} \frac{\left\langle  Q(t)^2 \right\rangle_T}{Vt}. 
\end{equation}
Here the bracket represents  
thermal average, and  
$Q(t)$ is the SU(2) topological charge, 
\begin{equation}
    Q(t) \equiv N_\mathrm{CS}(t) - N_\mathrm{CS}(0) = \frac{1}{32\pi^2} \int_0^t \mathrm{d}t' \int\mathrm{d}^3 x \epsilon_{\mu\nu\rho\sigma} \mathrm{Tr} W^{\mu\nu} W^{\rho\sigma}. 
\end{equation}
This determines the diffusion rate of the Chern-Simons number, including but not limited to the process described by a spharelon at $T<T_\mathrm{EW}$.
 In this way,
numerical calculation that evaluates Eq.~\eqref{diffrate} determines the full transition rate. 

Using recent lattice study~\cite{DOnofrio:2014rug} in the light of 
Higgs boson with a mass around 125 GeV, one can fit 
the sphaleron rate as
\begin{equation}
\label{lattice}
\begin{cases}
\Gamma_\mathrm{lattice,sph}(T)/T^4=(8.0\pm1.3)\times10^{-7} & (T>T_\mathrm{EW})\\
\textrm{log}(\Gamma_\mathrm{lattice,sph}(T)/T^4)=(0.83\pm0.01)T/\textrm{GeV}-(147.7\pm1.9) & (T<T_\mathrm{EW}), 
\end{cases}
\end{equation}
with $T_\mathrm{EW}=(159.5\pm1.5)$\,GeV.
We can also fit the temperature dependence of the Higgs VEV as
\begin{equation}
    \frac{v(T)}{T}\approx3\sqrt{1-\frac{T}{T_\mathrm{EW}}}. 
\end{equation}
The qualitative feature of these results can be explained 
by the analytical investigation
above;
see Eqs.~\eqref{sphaleronrateanal1} and \eqref{sphaleronrateanal2}. 
In other words, the lattice results quantitatively determine the unknown numerical coefficients in these expressions. 
In the following, we focus on the temperature regime below $T_\mathrm{EW}$. 
We incorporate transitions induced by configurations whose size is larger or smaller than the sphaleron solution introducing a size parameter, $a$, of the sphaleron-like configuration, motivated by
the argument of Ref.~\cite{Kharzeev:2019rsy}, and examine the transition rate through 
the configuration with each size.

\subsection{Sphaleron decoupling and baryogenesis}

\begin{figure}
    \centering
    \includegraphics{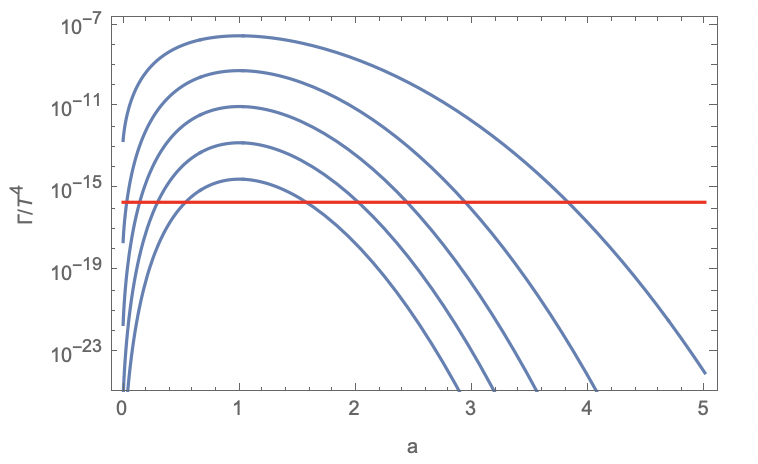}
    \caption{The sphaleron rate of different sizes and temperatures; 
see Eq.~\eqref{sizedepsphaleronrate}.
    The temperature is taken as $T=157.0,\,152.0,\,147.0,\,142.0,\,137.0\,$GeV from top to bottom, respectively. The red horizontal line represents the Hubble parameter normalized by the temperature, $H/T$, in the relevant temperature range. The temperature of the onset of the EW symmetry breaking is taken as $T_\mathrm{EW} = 160$ GeV. 
    and practically the integration range in Eq.~\eqref{sizedepsphaleronrate} is taken as $0.001 \leq a \leq 5$. }
    \label{sphaleronrate}
\end{figure}

Reference~\cite{Kharzeev:2019rsy} suggested that
the decoupling of sphaleron, which is 
an out-of-equilibrium process, can
generate BAU.
To investigate this possibility, we need to look at the process in depth.
The sphaleron process would proceed through not only the exact 
sphaleron solution (well fit by Eqs.~\eqref{sphaleronconff} and \eqref{sphaleronconfh}) but also higher energy field configurations
excited by thermal fluctuations.
While in the analytic calculation they are taken into account as
the prefactor of the exponential term in the transition 
rate evaluated by integrating over fluctuations
around the exact
sphaleron solution with the Gaussian approximation,
we here take into account the sphaleron-like configuration with different sizes, 
similar to the treatment of Ref.~\cite{Kharzeev:2019rsy}. 
That is, we express the sphaleron rate obtained 
by the lattice calculation~\cite{DOnofrio:2014rug} as
\begin{equation} \label{sizedepsphaleronrate}
    \Gamma_\mathrm{lattice,sph}(T)= \int \mathrm{d}a \, \Gamma_\mathrm{sph}(T,a), \quad \text{with} \quad \Gamma_\mathrm{sph}(T,a) = \frac{\mathrm{e}^{-M(T,a)/T}}{\int \mathrm{d}a'\, \mathrm{e}^{-M(T,a')/T}} \cdot \Gamma_\mathrm{lattice,sph}(T), 
\end{equation}
where $M(T,a)$ is the energy of a sphaleron-like configuration with the size $a$. 
In other words, we 
interpret that the results of the lattice simulation 
reflect the contributions from the field configuration with different sizes. 
Figure~\ref{sphaleronrate} shows the distribution of the 
sphaleron rate with respect to the size of the field configuration and temperatures below the EW scale, $T=157.0,\,152.0,\,147.0,\,142.0,\,137.0\,$GeV. 
It is peaked at $a=1$ corresponding to the true saddle point solution and becomes less than the 
Hubble scale for some smaller and larger values of $a$ at two points.\\ 
As the temperature decreases, the range of the sphaleron size above the horizontal line gets smaller, meaning that too large and too small sphalerons continuously decouple. Reference~\cite{Kharzeev:2019rsy} argued that this process is out of equilibrium and contributes to baryon number generation, because no upward-going fluctuations occur while decoupling and all configurations go down the energy barrier with these particular sizes of sphalerons. This kind of decoupling occurs below $T_\mathrm{EW}$ when the energy barrier gains a scale, and ends when all the sphaleron processes freeze. The authors of \cite{Kharzeev:2019rsy} used the sphaleron rate at decoupling, namely, when it coincides with the Hubble rate, to represent the rate of baryon generation. 

They also argued that the difference between the probability of production of one baryon number and that of reduction of one baryon number for sphaleron processes can be as large as
\begin{equation}
    A_\mathrm{CP}\sim0.25\times10^{-9}.
\end{equation}
Supposing that this process is active from the onset of the EW 
symmetry breaking to the sphaleron freezeout, $T_\mathrm{FO} \simeq 
130$ GeV (extrapolated from lattice simulation~\cite{DOnofrio:2014rug}), 
the resulting 
baryon asymmetry
has been estimated as
\begin{align}
\label{kestimate}
    \frac{n_B}{s}&=3A_\mathrm{CP}\times H_\mathrm{EW}\times(t_\mathrm{FO}-t_\mathrm{EW})/s_\mathrm{EW} \notag \\
    &    \sim 10^{-12}\left(\frac{A_\mathrm{CP}}{0.25 \times 10^{-9}}\right), 
\end{align}
where the factor 3 is the absolute value of baryon number produced from a sphaleron~\cite{Kharzeev:2019rsy}. 
The subscripts EW and FO represent that the variables 
are evaluated at the EW symmetry breaking and sphaleron freezeout,
respectively. 
This value is apparently just one order of magnitude smaller than 
the observed BAU, $n_B/s\sim 9 \times 10^{-11}$~\cite{Planck:2018nkj,ParticleDataGroup:2022pth}.
The authors of Ref.~\cite{Kharzeev:2019rsy} concluded that
this would be consistent with the present Universe
within the uncertainty of the estimate of the 
CP-violation $A_\mathrm{CP}$. 

\subsection{Sphaleron wash-out and kinetic equation}

Let us now point out that the estimate in Ref.~\cite{Kharzeev:2019rsy} did not include the effect from the sphalerons that are still in equilibrium, which tend to wash out the baryon number. 
In this subsection, we formulate the kinetic equation for the baryon asymmetry
that includes
this effect and also
the source term 
due to the sphaleron-like configurations
that are about to decouple. 
In the next section, we will evaluate it numerically.

At high enough temperature,
the sphaleron process tends to wash out the $B$+$L$ 
(the summation of baryon and lepton number) asymmetry~\cite{Kuzmin:1985mm}. 
When we do not take into account the baryon production from the 
sphaleron decoupling and any other source term, 
the kinetic equation for the baryon asymmetry 
around the EW scale 
is obtained 
as~\cite{Khlebnikov:1988sr}
\begin{equation}
 \frac{\mathrm{d}n_B}{\mathrm{d}t} + 3 H n_B = - \Gamma_B n_B, \quad \Gamma_B = \frac{39}{4} 
\frac{\Gamma_\mathrm{sph}(T)}{T^3}.    \label{kineticeq1}
\end{equation}
where we have assumed that there was  
no initial $B-L$ asymmetry.
Strictly speaking, the coefficient in $\Gamma_B$ depends on
the Higgs VEV and the equilibrium condition for the rapid processes 
such as the top Yukawa interaction~\cite{Khlebnikov:1988sr,Khlebnikov:1996vj}. 
But its variation during the process 
around $T\lesssim 10^2$ TeV
is quite small (a few percent)~\cite{Khlebnikov:1996vj,Laine:1999wv}
, and hence in 
the following, we take it as a constant, 39/4, 
which is often used in literature~e.g. \cite{Arnold:1987mh, Rubakov:1996vz}). 

We now rewrite 
the kinetic equation Eq.~\eqref{kineticeq1} for the baryogenesis scenario from the sphaleron
decoupling.
As discussed in the previous section, 
we introduce the source term from
decoupling sphalerons and modify the
wash-out term, too. 

To determine the source term
precisely, we first recall the sphaleron decoupling condition,
$$\frac{\Gamma_\mathrm{sph}(T)}{T^3}\lesssim H(T).$$
Since this condition is determined by the comparison between the Hubble 
expansion and the rate of the process, similar condition can be set for
each size of sphaleron-like configuration.
We assume  the decoupling temperature of the sphaleron-like configuration with 
its size $a$, $T_*(a)$, by the solution of
\begin{equation}
\label{Tstar}
\frac{\Gamma_\mathrm{sph} (T_*(a),a)}{T_*^3(a)}=cH(T_*(a)),
\end{equation}
where $c$ is a parameter of order of unity. Its precise value 
may be determined
by investigating the decoupling process quantitatively, which is left for
future research. 
We assume that when the rate $\Gamma_\mathrm{sph} (T_*(a),a)$ becomes  lower than $cT_*^3(a)H(T_*(a))$, the 
``size $a$ configuration'' decouples. 
In \cite{Kharzeev:2019rsy}, $c$ was set to $4/39$ to calculate $t_{FO}$ in (\ref{kestimate}) 
(see also Refs.~\cite{DOnofrio:2014rug,Burnier:2005hp}).
We will use it as a reference value.

With these criteria, one can modify the kinetic equation as follows. 
For given $c$ and a temperature $T$, we can draw a horizontal line 
for $cH(T)/T$ in Figure~\ref{sphaleronrate} 
(the red line corresponds to the case with $c=1$).
At each temperature, $T$, we identify the points where the curve of $\Gamma_\mathrm{sph}(T,a)$ crosses the horizontal line $cH(T_*)/T_*$, which 
we call $a=a_l$ and $a=a_u$ with $a_l\leq a_u$. 
Sphaleron-like configurations with size $a_l\leq a \leq a_u$
contribute to washing out the asymmetry
while those with $a \leq a_l$ and $a \geq a_u$ may
produce net baryon number. 
As the temperature drops, the curve will fall below the horizontal line. 
We define this critical temperature as $T_c$.
Then taking into account the factor for the efficiency of 
baryon-number production, $3 \times A_\mathrm{CP}$,
we determine the source term for the baryon-number generation, $P(T)$, as
\begin{equation}
    P(T)=
\begin{cases}
\left(\int^{a_l}_{a_\mathrm{min}}\mathrm{d}a\ \Gamma_\mathrm{sph} (T,a) +\int^{a_\mathrm{max}}_{a_u} \mathrm{d}a\ \Gamma_\mathrm{sph} (T,a)\right)\cdot 3A_\mathrm{CP}, & \text{for} \quad T_c< T\leq T_\mathrm{EW}, \\
\Gamma_\mathrm{lattice,sph}\cdot 3A_\mathrm{CP}, & \text{for} \quad T\leq T_c,
\end{cases}
\label{source}
\end{equation}
and the wash-out rate as
\begin{equation}
 \Gamma_B(T) =  \frac{39}{4T^3} \,  \Gamma_\mathrm{washout,sph}(T)=
\begin{cases}
 \dfrac{39}{4T^3} \int_{a_l}^{a_u}\mathrm{d}a\ \Gamma_\mathrm{sph} (T,a),  & \text{for} \quad T_c<T\leq T_\mathrm{EW}, \\
0, & \text{for} \quad T\leq T_c.
\end{cases}
\label{washout}
\end{equation}
For the practical purpose, in calculating $\Gamma_\mathrm{sph}(T,a)$ and $P(T)$, 
we choose $a_\mathrm{min} =0.001$ and $a_\mathrm{max}=5$. If we do not find $a_l$ for $P(T)$, we just omit the first term in Eq.~\eqref{source}. 
The contributions from the omitted part is exponentially small, 
and hence the results are unaffected. 
\noindent The main contribution of the source term calculated in this way comes from the sphalerons near the horizontal red line
as we can see
in Figure~\ref{sphaleronrate} where the vertical axis is of logarithmic scale. 

By taking them together, the modified kinetic equation for our scenario is 
given as
\begin{equation}
-HT\frac{\mathrm{d}n_B}{\mathrm{d}T}+3Hn_B=- \frac{39}{4}\frac{\Gamma_\mathrm{washout,sph}(T)}{T^3}n_B+P(T), 
\label{kineticeq}
\end{equation}
where we have used the relation $H=1/2t \propto T^2$ during the radiation dominated era.

\section{Resulting baryon-to-entropy ratio} \label{sec:result}

\begin{figure}
  \begin{minipage}[b]{0.45\linewidth}
    \centering  \includegraphics[width=9cm]{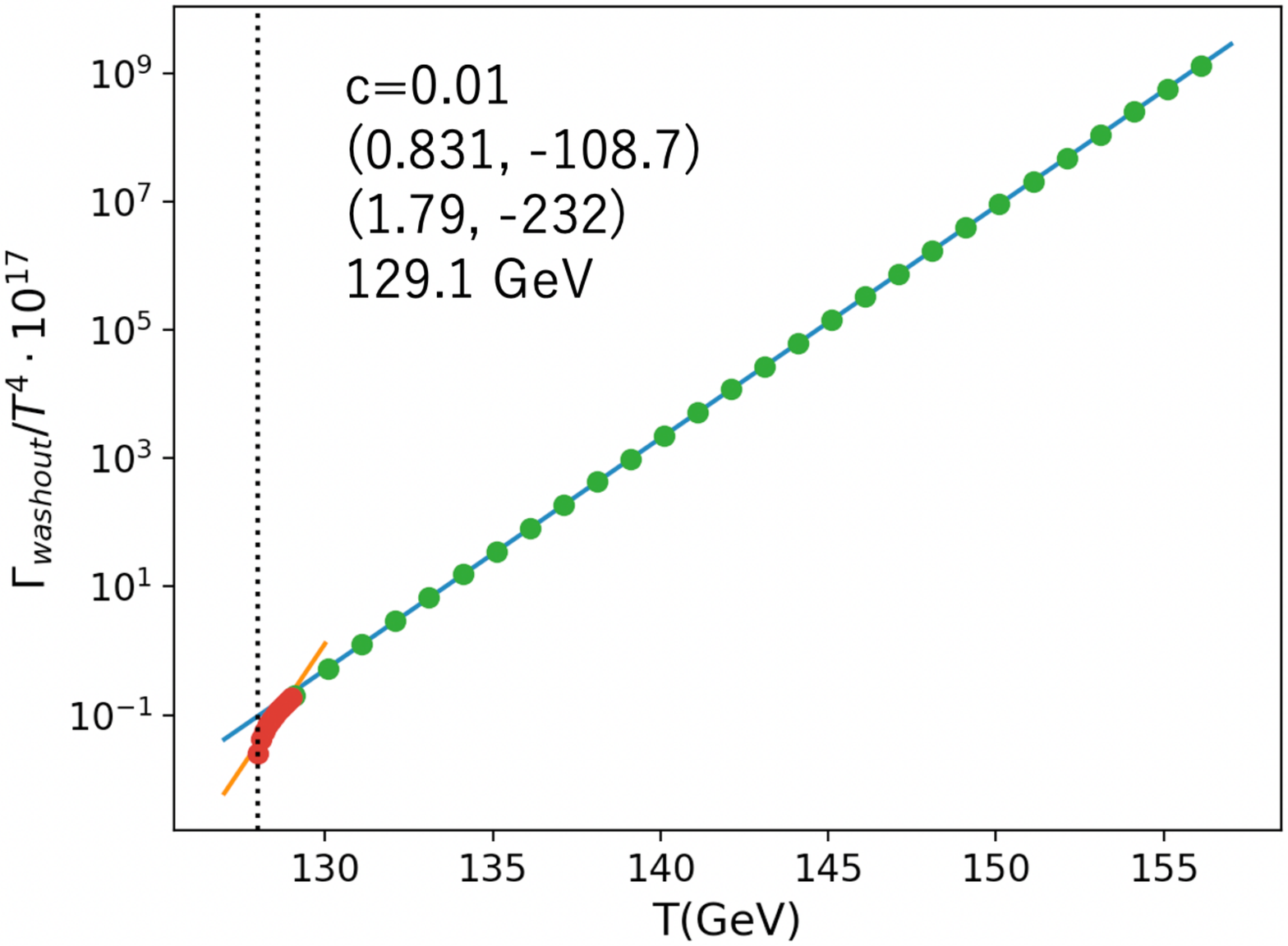}
  \end{minipage}
  \begin{minipage}[b]{0.45\linewidth}
    \centering  \includegraphics[width=9cm]{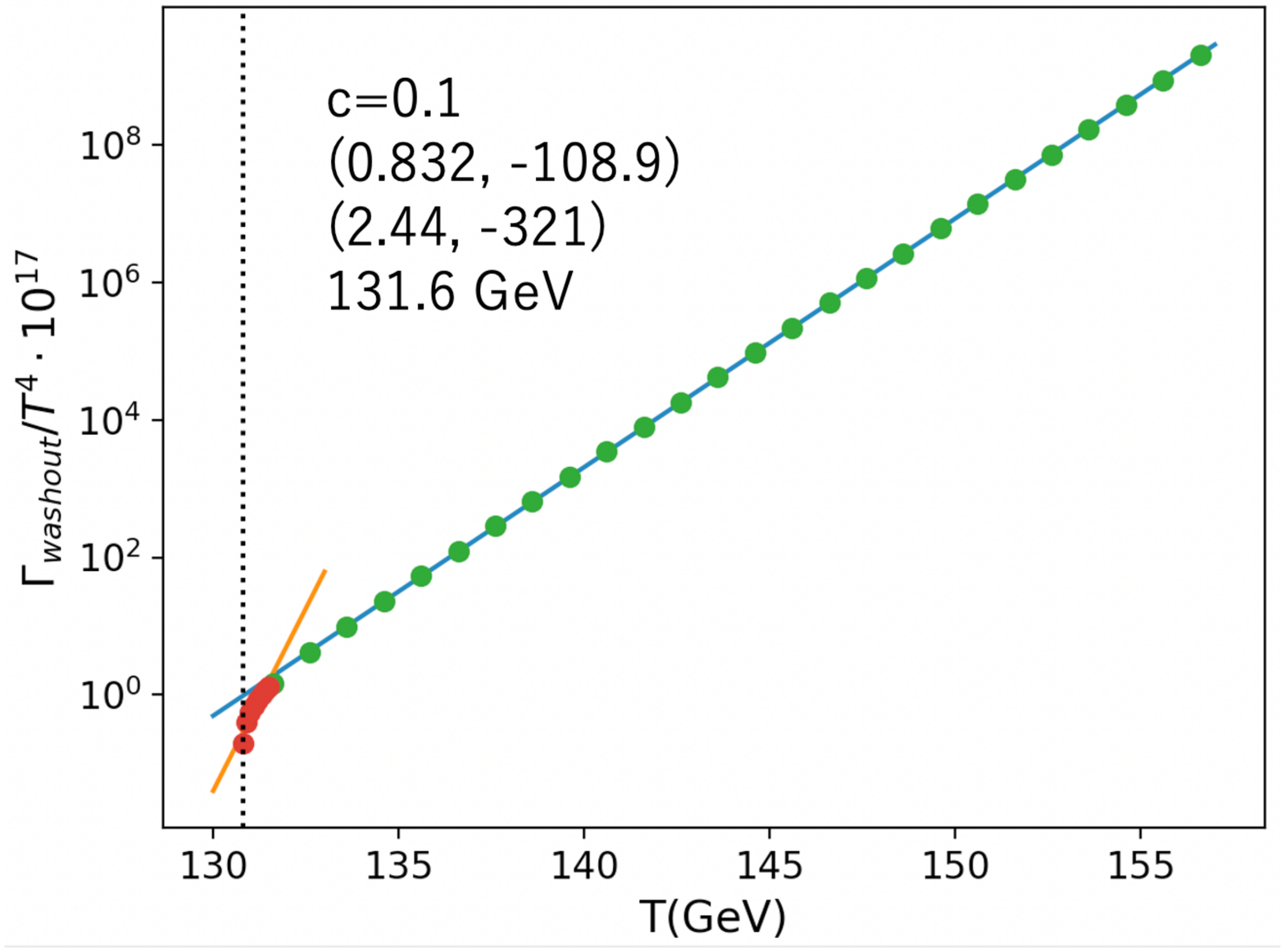}
  \end{minipage}\\
    \begin{minipage}[b]{0.45\linewidth}
    \centering  \includegraphics[width=9cm]{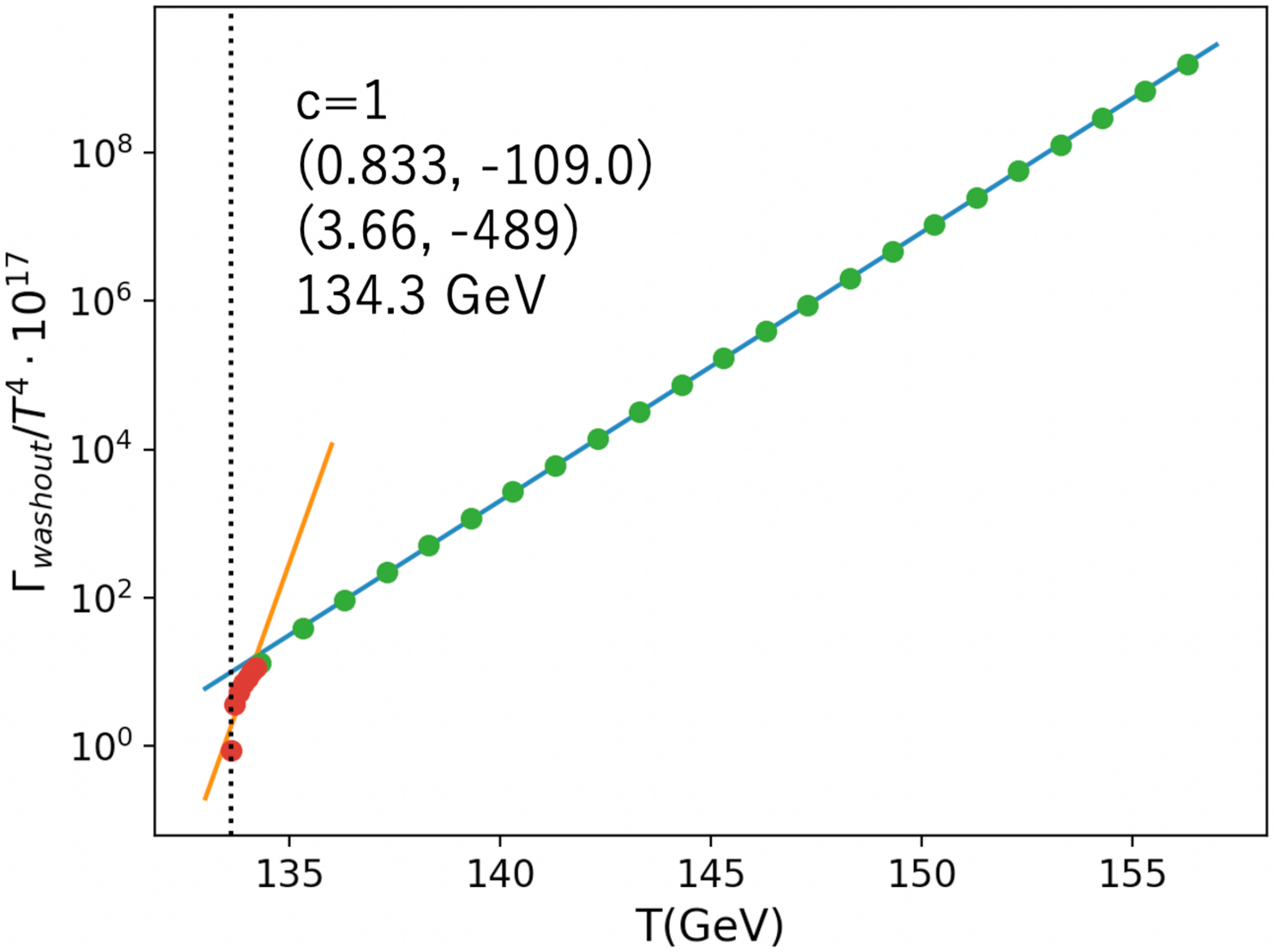}
  \end{minipage}  
    \caption{
    Numerically evaluated values of the wash-out term,
    $\Gamma_\mathrm{washout,sph}/T^4$, 
    (see Eq.~\eqref{washout}) as well as its
    fitting curve (Eq.~\eqref{fittingws}) for $c=0.01, 0.1$ and 1.
    The numbers in each figure are the parameters in the fitting curve:  from top to bottom, ($n_1$, $n_2$), ($n_3$, $n_4$), and $n_5\,$GeV. 
    The vertical dotted lines represent $T=T_c$. 
    }
    \label{fittingwash}
\end{figure}

\begin{figure}
  \begin{minipage}[b]{0.45\linewidth}
    \centering  \includegraphics[width=9cm]{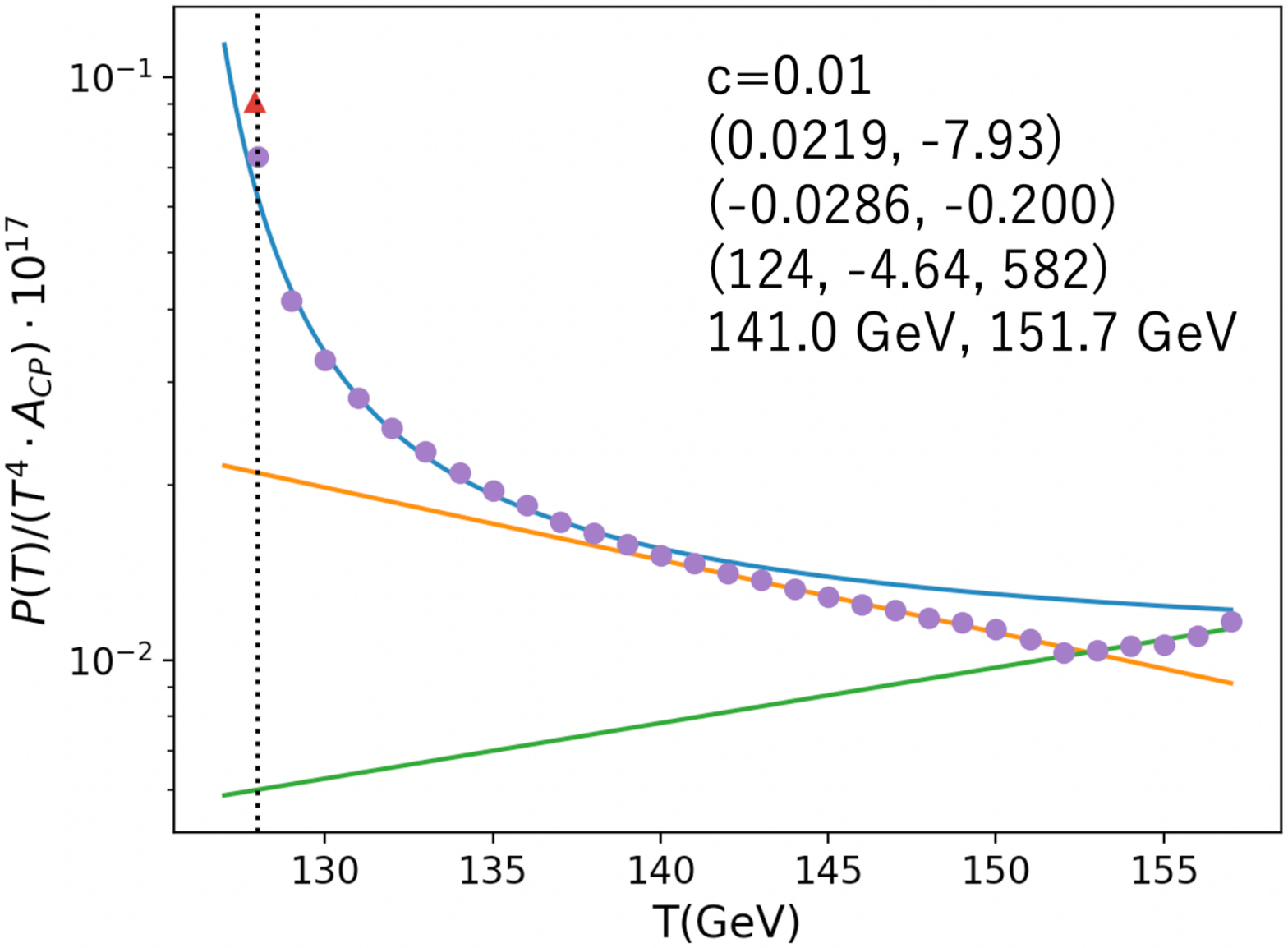}
  \end{minipage}
  \begin{minipage}[b]{0.45\linewidth}
    \centering  \includegraphics[width=9cm]{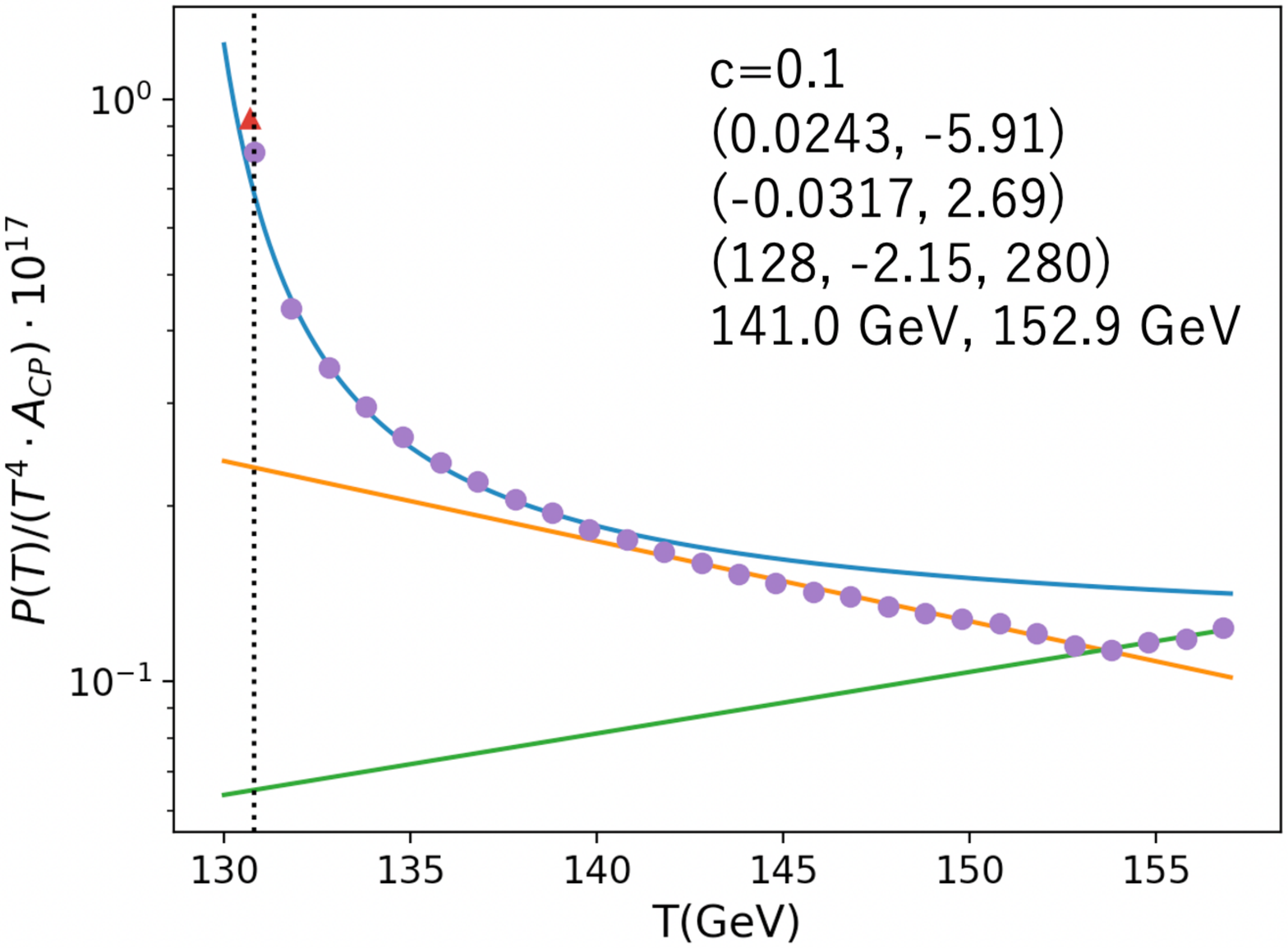}
  \end{minipage}\\
    \begin{minipage}[b]{0.45\linewidth}
    \centering  \includegraphics[width=9cm]{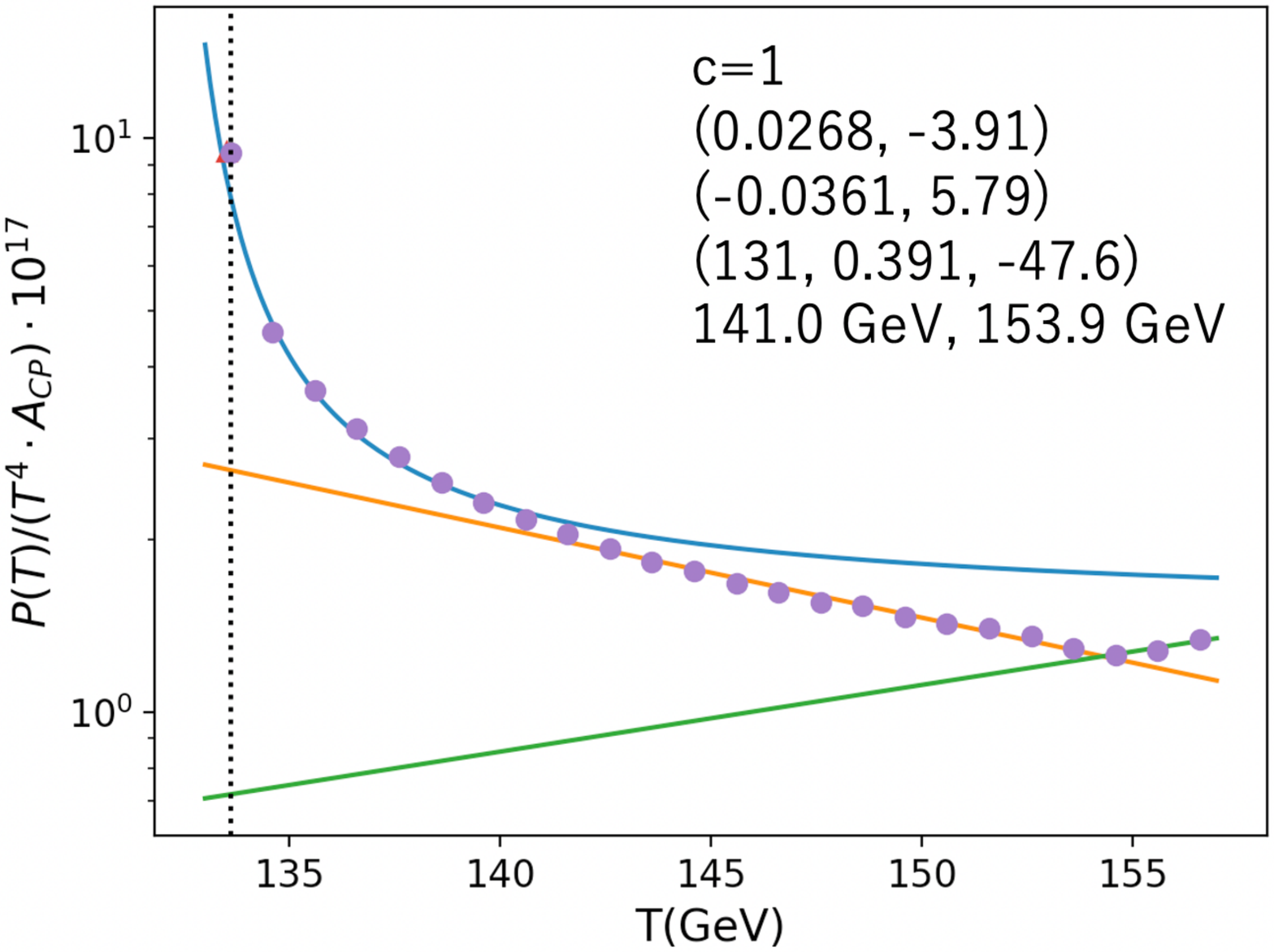}
  \end{minipage}
    \caption{
    Numerically evaluated values of the source term,
    $P/(T^4 \cdot A_\mathrm{CP})$, 
    (see Eq.~\eqref{source}) as well as its
    fitting curve (Eq.~\eqref{fittingp}). 
    The vertical dotted line represent $T=T_c$.
    The red triangle denotes the value obtained from the lattice result
    (Eq.~\eqref{lattice})
    at $(T_c-0.1\,\textrm{GeV})$. 
    The numbers  in each figure are the parameters in the fitting curve: from top to bottom, ($m_1$, $m_2$), ($m_3$, $m_4$), ($m_5$, $m_6$, $m_7$), and $m_8\,$GeV, $m_9\,$GeV. 
    }
    \label{fittingsource}
\end{figure}

Now we solve the kinetic equation numerically 
to determine the resultant baryon asymmetry. 
For this purpose, 
first we evaluate the 
source term and wash-out term. 
For temperatures in the range $125.0 \, \mathrm{GeV} \leq T \leq 157.0 \, \mathrm{GeV}$
every 0.1 GeV, 
we evaluate $\Gamma_\mathrm{sph} (T,a)$ for $0.001<a<5$ 
by using Eqs.~\eqref{sphaleronconff}, \eqref{sphaleronconfh}, \eqref{sphaleronmass}, \eqref{lattice}, and \eqref{sizedepsphaleronrate}. 
Then for a given $c$, we can determine $a_l$ and $a_u$ 
as a function of $T$. 
We first determine the net decoupling temperature $T_*$ 
by the condition $\Gamma_\mathrm{lattice}(T_*)/T_*^4 = c H(T_*)/T_*$, 
then determine $a_l$ and $a_u$ 
by the solution of 
$\Gamma_\mathrm{sph} (T,a_{l,u}(T))/T^4 = c H(T_*)/T_*$,  
which makes it possible to determine $T_c$ as a solution of $\mathrm{max} [\Gamma(T_c,a)/T_c^4] = c H(T_*)/T_*$.  
With these $a_l(T)$ and $a_u(T)$ we calculate $P(T)$ and 
$\Gamma_\mathrm{washout, sph}(T)$ by using Eqs.~\eqref{source} and \eqref{washout}.
We find that the wash-out term is well fit as
\begin{equation} \label{fittingws}
    \Gamma_\mathrm{washout,sph}(T) = 10^{-17}\, T^4 \times \begin{cases} 
    \exp \left(n_1 T + n_2\right), & \text{for}
    \quad n_5 \,\mathrm{GeV}< T< T_\mathrm{EW}, \\
    \exp \left(n_3 T + n_4 \right), & \text{for} \quad T_c< T< n_5 \,\mathrm{GeV}, 
    \end{cases}
\end{equation}
while the source term is well fit as
\begin{equation} \label{fittingp}
    P(T) = 10^{-17} A_\mathrm{CP} T^4 \times \begin{cases} 
    \exp \left(m_1 T + m_2\right), & \text{for} \quad m_9 \,\mathrm{GeV}< T< T_\mathrm{EW}\, \mathrm{GeV}, \\
    \exp \left(m_3 T + m_4\right), & \text{for} \quad m_8 \,\mathrm{GeV} < T< m_9 \,\mathrm{GeV},\\
    \exp \left[(m_6 T + m_7)/(T-m_5) \right], & \text{for} \quad T_c< T< m_8 \,\mathrm{GeV}    
    \end{cases}
\end{equation}
where $n_1$ to $n_4$ as well as $m_1$ to $m_9$ depend on $c$. 
Figures \ref{fittingwash} and \ref{fittingsource} show $ \Gamma_\mathrm{washout,sph}(T)$ and $P(T)$ 
numerically obtained for $c=0.01, 0.1$ and $1$, respectively, 
which clearly show the fitting functions Eqs.~\eqref{fittingws} and 
\eqref{fittingp} work very well.
Note that we find $T_*= 128.7, 131.5, 134.3$ GeV and $T_c= 128.0, 130.8, 133.6$ GeV, for $c=0.01, 0.1, 1$, 
respectively.

\begin{figure}
    \centering
    \includegraphics[width=15cm]{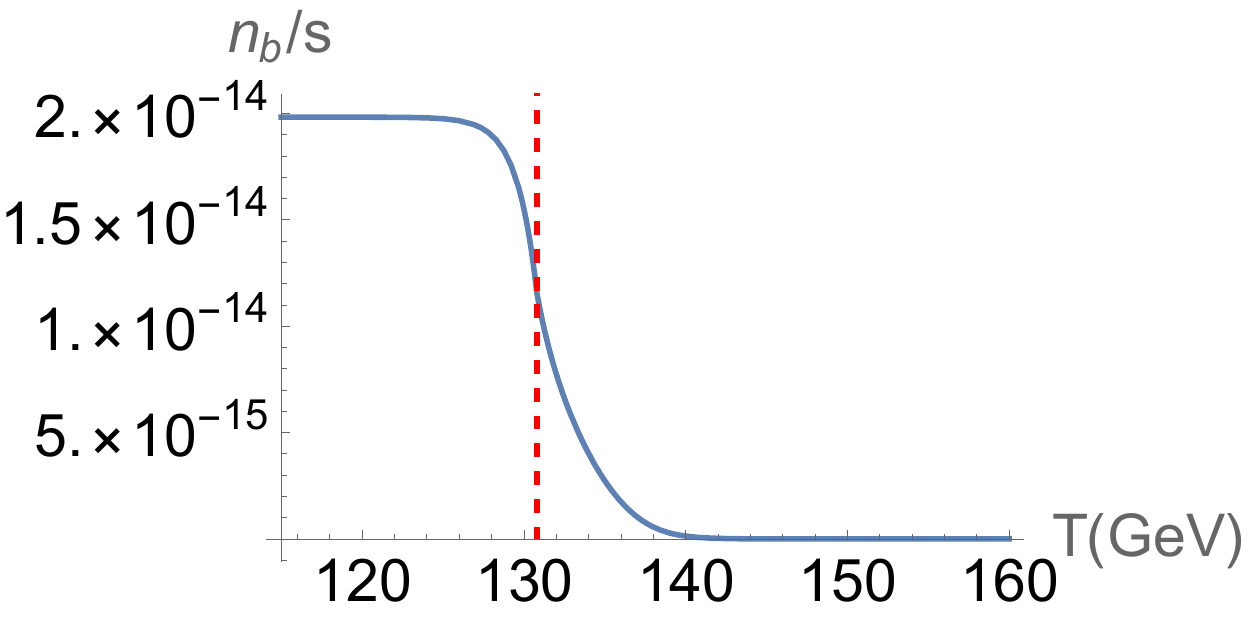}
    \caption{The time evolution of the baryon-to-entropy ratio 
    in the case $c=0.1$. The red dashed line represents $T=T_c=130.8$ GeV.}
    \label{baryonevolve}
\end{figure}

Putting these fitting results, we solve the kinetic equation (\ref{kineticeq}) 
for $T<T_\mathrm{EW} = 160$ GeV  taking $n_B=0$ as the initial condition. 
Figure~\ref{baryonevolve} shows the typical evolution of baryon asymmetry for $c=0.1$. 
We can see that baryon asymmetry stops 
growing
some temperature below $T_c = 130.8$ GeV and
becomes constant
at $T\lesssim 125$ GeV. 

\begin{figure}
    \centering
    \includegraphics[width=15cm]{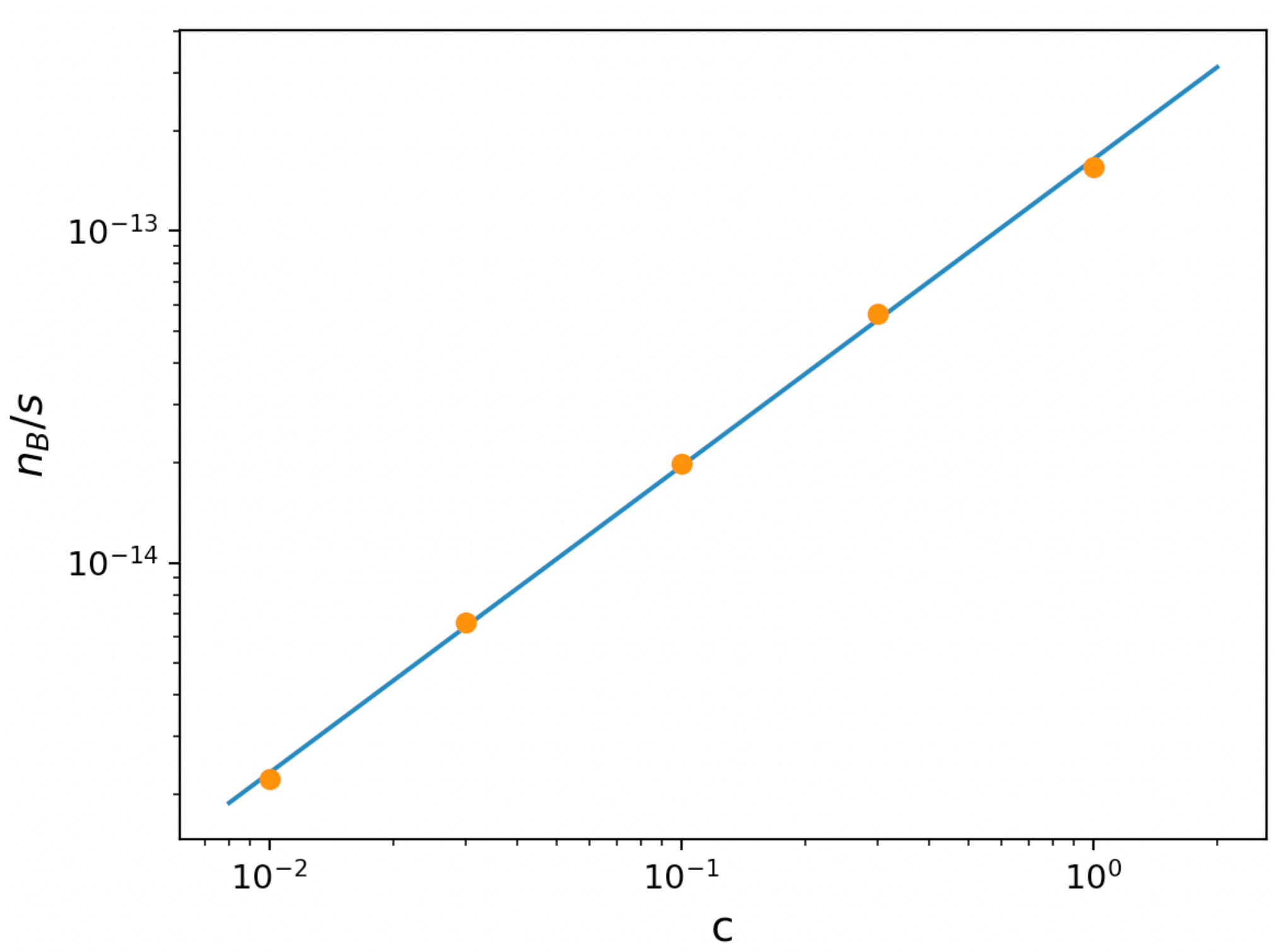}
    \caption{The net baryon-to-entropy ratioevaluated at $T=120$ GeV
    for $c=0.01, 0.03, 0.1, 0.3$ and 1, as well as its fitting function
    (Eq.~\eqref{fittingbaryon}).
    Here we have taken $A_\mathrm{CP}= 0.25 \times 10^{-9}$.} 
    \label{baryonn1}
\end{figure}

The resulting baryon asymmetry at 120\,GeV, when it has already become a constant,
is plotted as a function of $c$ in Figure~\ref{baryonn1}. Here we have set $A_\mathrm{CP} = 0.25 \times 10^{-9}$~\cite{Kharzeev:2019rsy} and  
used $s=\frac{4\pi^2}{90}g_*T^3$ for the process calculated above to convert $T^3$ to $s$. The baryon-to-entropy ratio increases as $c$ increases as expected.
Considering that the result is proportional to the
CP violation parameter $A_\mathrm{CP}$, which has not been calculated quantitatively, we give a general expression of our conclusion:
\begin{equation} \label{fittingbaryon}
    \frac{n_B}{s}\sim 3 \times 10^{-14} \left(\frac{c}{4/39}\right)^{0.92} \frac{A_\mathrm{CP}}{0.25\times10^{-9}}.
\end{equation}
 
Comparing with the observed value of baryon-to-entropy ratio is \cite{Planck:2018nkj}
\begin{equation}
    \frac{n_B}{s}\sim9\times10^{-11}, 
\end{equation}
the baryon-to-entropy ratio in our calculation is two to three orders lower than the observed value. 
We conclude that, unfortunately, as long as we take the parameter  $c$ 
of order of unity, this scenario is impossible to explain the present BAU, 
even with an optimistic estimate of the CP-violation. On the other hand,
we expect that the present result also qualitatively 
applies to extensions of the SM, as long as the temperature scale of electroweak symmetry-breaking and Higgs VEV does not differ dramatically from what we studied here.
Therefore, if some new physics enhances the CP-violation up to 
$A_\mathrm{CP} \sim 10^{-6}$, the present BAU might be explained. 
Compared with the standard EW baryogenesis~\cite{Morrissey:2012db}, 
it is more economical since we do not have to make the EW symmetry breaking a strong first-order PT.

One might be concerned that when the sphaleron size is too small, quantum fluctuations become large and the classical field configuration loses sense.
Typical amplitudes of fluctuations of $h(\xi)$ and $f(\xi)$ can be estimated as \cite{Mukhanov:2007}
\begin{equation}
    |\delta h|\simeq\frac{\sigma_h}{\Omega},
\end{equation}
\begin{equation}
    |\delta f|\simeq\frac{\sigma_f}{\Xi},
\end{equation}
where $\sigma_h$ and $\sigma_f$ are constants of order unity. 
This suggests that the notion of the sphaleron-like configuration with $a<0.1$
might be questionable.
To check the effect of possible non-contributions from these parameters,
we took $a\in[0.1,5]$ and repeated the procedures above. We  
have found that the result does not change, because sphaleron-like configurations with 
$a<0.1$ do not contribute much.
Therefore, our conclusion is unchanged even if we take into account this concern. 

\section{Conclusion} \label{sec:conclusion}
Reference \cite{Kharzeev:2019rsy} proposed a scenario of baryogenesis within the SM, where the decoupling of sphalerons provides the inequilibrium process required 
by Sakharov's conditions. They focused on
the size distribution of sphalerons, 
and pointed out that sphalerons of different sizes and rates keep decoupling from the moment of EWPT to that 
of freeze-out of the entire sphaleron process.
Sphaleron decoupling described in such a way 
can provide a source for the
generation of baryon number. 
With an estimate of
the difference between probability of production of quarks and antiquarks in sphaleron process, or the CP-violation, $A_\mathrm{CP}$,
they gave an estimate
of $n_B/s$, which was only one order smaller than the observed BAU. 
They argued that the BAU can be explained by the SM taking into account
the uncertainty in their crude estimation of $A_\mathrm{CP}$.

In this paper, we have studied the scenario following the idea of Ref.~\cite{Kharzeev:2019rsy} with constructing 
a more proper kinetic equation
for the BAU. We have taken into account
the wash-out effect from the sphalerons that are still in equilibrium, which was not 
incorporated in Ref.~\cite{Kharzeev:2019rsy}.  We have calculated the rate distribution for sphalerons of different sizes at different temperatures, and evaluated the source term and wash-out term in the kinetic equation. Including this effect, we have estimated 
$n_B/s$ numerically, which turned out to be
two to three orders smaller than the observed BAU, 
with the amounts of CP violation used in \cite{Kharzeev:2019rsy}. 
Thus we conclude the scenario of baryogenesis in the SM proposed in \cite{Kharzeev:2019rsy} does not work unless 
$A_\mathrm{CP}$ is enhanced by a factor of $10$ or $10^2$ 
compared with their optimistic value.
In other words, in order to make use of the sphaleron decoupling discussed here to realize adequate baryogenesis, 
we must introduce some new ingredients to 
increase the amount of 
CP violation to the level of $A_\mathrm{CP}\sim 10^{-6}$.

\bibliographystyle{utphys}
\bibliography{ref}

\providecommand{\href}[2]{#2}\begingroup\raggedright\begin{thebibliography}{10}

\bibitem{Sakharov:1967dj}
A.~D. Sakharov, ``{Violation of CP Invariance, C asymmetry, and baryon
  asymmetry of the universe},''
  \href{http://dx.doi.org/10.1070/PU1991v034n05ABEH002497}{{\em Pisma Zh. Eksp.
  Teor. Fiz.} {\bfseries 5} (1967) 32--35}.

\bibitem{Kuzmin:1985mm}
V.~A. Kuzmin, V.~A. Rubakov, and M.~E. Shaposhnikov, ``{On the Anomalous
  Electroweak Baryon Number Nonconservation in the Early Universe},''
  \href{http://dx.doi.org/10.1016/0370-2693(85)91028-7}{{\em Phys. Lett. B}
  {\bfseries 155} (1985) 36}.

\bibitem{Fukugita:1986hr}
M.~Fukugita and T.~Yanagida, ``{Baryogenesis Without Grand Unification},''
  \href{http://dx.doi.org/10.1016/0370-2693(86)91126-3}{{\em Phys. Lett. B}
  {\bfseries 174} (1986) 45--47}.

\bibitem{Affleck:1984fy}
I.~Affleck and M.~Dine, ``{A New Mechanism for Baryogenesis},''
  \href{http://dx.doi.org/10.1016/0550-3213(85)90021-5}{{\em Nucl. Phys. B}
  {\bfseries 249} (1985) 361--380}.

\bibitem{Bodeker:2020ghk}
D.~Bodeker and W.~Buchmuller, ``{Baryogenesis from the weak scale to the grand
  unification scale},''
  \href{http://dx.doi.org/10.1103/RevModPhys.93.035004}{{\em Rev. Mod. Phys.}
  {\bfseries 93} no.~3, (2021) 035004},
  \href{http://arxiv.org/abs/2009.07294}{{\ttfamily arXiv:2009.07294
  [hep-ph]}}.

\bibitem{Planck:2018nkj}
{\bfseries Planck} Collaboration, N.~Aghanim {\em et~al.}, ``{Planck 2018
  results. I. Overview and the cosmological legacy of Planck},''
  \href{http://dx.doi.org/10.1051/0004-6361/201833880}{{\em Astron. Astrophys.}
  {\bfseries 641} (2020) A1}, \href{http://arxiv.org/abs/1807.06205}{{\ttfamily
  arXiv:1807.06205 [astro-ph.CO]}}.

\bibitem{ParticleDataGroup:2022pth}
{\bfseries Particle Data Group} Collaboration, R.~L. Workman {\em et~al.},
  ``{Review of Particle Physics},''
  \href{http://dx.doi.org/10.1093/ptep/ptac097}{{\em PTEP} {\bfseries 2022}
  (2022) 083C01}.

\bibitem{DOnofrio:2015gop}
M.~D'Onofrio and K.~Rummukainen, ``{Standard model cross-over on the
  lattice},'' \href{http://dx.doi.org/10.1103/PhysRevD.93.025003}{{\em Phys.
  Rev. D} {\bfseries 93} no.~2, (2016) 025003},
  \href{http://arxiv.org/abs/1508.07161}{{\ttfamily arXiv:1508.07161
  [hep-ph]}}.

\bibitem{Buchmuller:1994qy}
W.~Buchmuller and O.~Philipsen, ``{Phase structure and phase transition of the
  SU(2) Higgs model in three-dimensions},''
  \href{http://dx.doi.org/10.1016/0550-3213(95)00124-B}{{\em Nucl. Phys. B}
  {\bfseries 443} (1995) 47--69},
  \href{http://arxiv.org/abs/hep-ph/9411334}{{\ttfamily arXiv:hep-ph/9411334}}.

\bibitem{Kajantie:1996mn}
K.~Kajantie, M.~Laine, K.~Rummukainen, and M.~E. Shaposhnikov, ``{Is there a~
  hot electroweak phase transition at $m_H \gtrsim m_W$?},''
  \href{http://dx.doi.org/10.1103/PhysRevLett.77.2887}{{\em Phys. Rev. Lett.}
  {\bfseries 77} (1996) 2887--2890},
  \href{http://arxiv.org/abs/hep-ph/9605288}{{\ttfamily arXiv:hep-ph/9605288}}.

\bibitem{Csikor:1998eu}
F.~Csikor, Z.~Fodor, and J.~Heitger, ``{Endpoint of the hot electroweak phase
  transition},'' \href{http://dx.doi.org/10.1103/PhysRevLett.82.21}{{\em Phys.
  Rev. Lett.} {\bfseries 82} (1999) 21--24},
  \href{http://arxiv.org/abs/hep-ph/9809291}{{\ttfamily arXiv:hep-ph/9809291}}.

\bibitem{ATLAS:2012yve}
{\bfseries ATLAS} Collaboration, G.~Aad {\em et~al.}, ``{Observation of a new
  particle in the search for the Standard Model Higgs boson with the ATLAS
  detector at the LHC},''
  \href{http://dx.doi.org/10.1016/j.physletb.2012.08.020}{{\em Phys. Lett. B}
  {\bfseries 716} (2012) 1--29},
  \href{http://arxiv.org/abs/1207.7214}{{\ttfamily arXiv:1207.7214 [hep-ex]}}.

\bibitem{CMS:2012qbp}
{\bfseries CMS} Collaboration, S.~Chatrchyan {\em et~al.}, ``{Observation of a
  New Boson at a Mass of 125 GeV with the CMS Experiment at the LHC},''
  \href{http://dx.doi.org/10.1016/j.physletb.2012.08.021}{{\em Phys. Lett. B}
  {\bfseries 716} (2012) 30--61},
  \href{http://arxiv.org/abs/1207.7235}{{\ttfamily arXiv:1207.7235 [hep-ex]}}.

\bibitem{Akhmedov:1998qx}
E.~K. Akhmedov, V.~A. Rubakov, and A.~Y. Smirnov, ``{Baryogenesis via neutrino
  oscillations},'' \href{http://dx.doi.org/10.1103/PhysRevLett.81.1359}{{\em
  Phys. Rev. Lett.} {\bfseries 81} (1998) 1359--1362},
  \href{http://arxiv.org/abs/hep-ph/9803255}{{\ttfamily arXiv:hep-ph/9803255}}.

\bibitem{Asaka:2005pn}
T.~Asaka and M.~Shaposhnikov, ``{The $\nu$MSM, dark matter and baryon asymmetry
  of the universe},''
  \href{http://dx.doi.org/10.1016/j.physletb.2005.06.020}{{\em Phys. Lett. B}
  {\bfseries 620} (2005) 17--26},
  \href{http://arxiv.org/abs/hep-ph/0505013}{{\ttfamily arXiv:hep-ph/0505013}}.

\bibitem{Manton:1983nd}
N.~S. Manton, ``{Topology in the Weinberg-Salam Theory},''
  \href{http://dx.doi.org/10.1103/PhysRevD.28.2019}{{\em Phys. Rev. D}
  {\bfseries 28} (1983) 2019}.

\bibitem{Klinkhamer:1984di}
F.~R. Klinkhamer and N.~S. Manton, ``{A Saddle Point Solution in the
  Weinberg-Salam Theory},''
  \href{http://dx.doi.org/10.1103/PhysRevD.30.2212}{{\em Phys. Rev. D}
  {\bfseries 30} (1984) 2212}.

\bibitem{Arnold:1987mh}
P.~B. Arnold and L.~D. McLerran, ``{Sphalerons, Small Fluctuations and Baryon
  Number Violation in Electroweak Theory},''
  \href{http://dx.doi.org/10.1103/PhysRevD.36.581}{{\em Phys. Rev. D}
  {\bfseries 36} (1987) 581}.

\bibitem{tHooft:1976rip}
G.~'t~Hooft, ``{Symmetry Breaking Through Bell-Jackiw Anomalies},''
  \href{http://dx.doi.org/10.1103/PhysRevLett.37.8}{{\em Phys. Rev. Lett.}
  {\bfseries 37} (1976) 8--11}.

\bibitem{Adler:1969gk}
S.~L. Adler, ``{Axial vector vertex in spinor electrodynamics},''
  \href{http://dx.doi.org/10.1103/PhysRev.177.2426}{{\em Phys. Rev.} {\bfseries
  177} (1969) 2426--2438}.

\bibitem{Bell:1969ts}
J.~S. Bell and R.~Jackiw, ``{A PCAC puzzle: $\pi^0 \to \gamma \gamma$ in the
  $\sigma$ model},'' \href{http://dx.doi.org/10.1007/BF02823296}{{\em Nuovo
  Cim. A} {\bfseries 60} (1969) 47--61}.

\bibitem{Jackiw:1976pf}
R.~Jackiw and C.~Rebbi, ``{Vacuum Periodicity in a Yang-Mills Quantum
  Theory},'' \href{http://dx.doi.org/10.1103/PhysRevLett.37.172}{{\em Phys.
  Rev. Lett.} {\bfseries 37} (1976) 172--175}.

\bibitem{Shaposhnikov:1987tw}
M.~E. Shaposhnikov, ``{Baryon Asymmetry of the Universe in Standard Electroweak
  Theory},'' \href{http://dx.doi.org/10.1016/0550-3213(87)90127-1}{{\em Nucl.
  Phys. B} {\bfseries 287} (1987) 757--775}.

\bibitem{Farrar:1993sp}
G.~R. Farrar and M.~E. Shaposhnikov, ``{Baryon asymmetry of the universe in the
  minimal Standard Model},''
  \href{http://dx.doi.org/10.1103/PhysRevLett.70.2833}{{\em Phys. Rev. Lett.}
  {\bfseries 70} (1993) 2833--2836},
  \href{http://arxiv.org/abs/hep-ph/9305274}{{\ttfamily arXiv:hep-ph/9305274}}.
  [Erratum: Phys.Rev.Lett. 71, 210 (1993)].

\bibitem{Farrar:1993hn}
G.~R. Farrar and M.~E. Shaposhnikov, ``{Baryon asymmetry of the universe in the
  standard electroweak theory},''
  \href{http://dx.doi.org/10.1103/PhysRevD.50.774}{{\em Phys. Rev. D}
  {\bfseries 50} (1994) 774},
  \href{http://arxiv.org/abs/hep-ph/9305275}{{\ttfamily arXiv:hep-ph/9305275}}.

\bibitem{Gavela:1993ts}
M.~B. Gavela, P.~Hernandez, J.~Orloff, and O.~Pene, ``{Standard model CP
  violation and baryon asymmetry},''
  \href{http://dx.doi.org/10.1142/S0217732394000629}{{\em Mod. Phys. Lett. A}
  {\bfseries 9} (1994) 795--810},
  \href{http://arxiv.org/abs/hep-ph/9312215}{{\ttfamily arXiv:hep-ph/9312215}}.

\bibitem{Gavela:1994dt}
M.~B. Gavela, P.~Hernandez, J.~Orloff, O.~Pene, and C.~Quimbay, ``{Standard
  model CP violation and baryon asymmetry. Part 2: Finite temperature},''
  \href{http://dx.doi.org/10.1016/0550-3213(94)00410-2}{{\em Nucl. Phys. B}
  {\bfseries 430} (1994) 382--426},
  \href{http://arxiv.org/abs/hep-ph/9406289}{{\ttfamily arXiv:hep-ph/9406289}}.

\bibitem{Kharzeev:2019rsy}
D.~Kharzeev, E.~Shuryak, and I.~Zahed, ``{Sphalerons, baryogenesis, and helical
  magnetogenesis in the electroweak transition of the minimal standard
  model},'' \href{http://dx.doi.org/10.1103/PhysRevD.102.073003}{{\em Phys.
  Rev. D} {\bfseries 102} no.~7, (2020) 073003},
  \href{http://arxiv.org/abs/1906.04080}{{\ttfamily arXiv:1906.04080
  [hep-ph]}}.

\bibitem{Kajantie:1995dw}
K.~Kajantie, M.~Laine, K.~Rummukainen, and M.~E. Shaposhnikov, ``{Generic rules
  for high temperature dimensional reduction and their application to the
  standard model},'' \href{http://dx.doi.org/10.1016/0550-3213(95)00549-8}{{\em
  Nucl. Phys. B} {\bfseries 458} (1996) 90--136},
  \href{http://arxiv.org/abs/hep-ph/9508379}{{\ttfamily arXiv:hep-ph/9508379}}.

\bibitem{Arnold:1996dy}
P.~B. Arnold, D.~Son, and L.~G. Yaffe, ``{The Hot baryon violation rate is O
  (alpha-w**5 T**4)},'' \href{http://dx.doi.org/10.1103/PhysRevD.55.6264}{{\em
  Phys. Rev. D} {\bfseries 55} (1997) 6264--6273},
  \href{http://arxiv.org/abs/hep-ph/9609481}{{\ttfamily arXiv:hep-ph/9609481}}.

\bibitem{DOnofrio:2014rug}
M.~D'Onofrio, K.~Rummukainen, and A.~Tranberg, ``{Sphaleron Rate in the Minimal
  Standard Model},''
  \href{http://dx.doi.org/10.1103/PhysRevLett.113.141602}{{\em Phys. Rev.
  Lett.} {\bfseries 113} no.~14, (2014) 141602},
  \href{http://arxiv.org/abs/1404.3565}{{\ttfamily arXiv:1404.3565 [hep-ph]}}.

\bibitem{Khlebnikov:1988sr}
S.~Y. Khlebnikov and M.~E. Shaposhnikov, ``{The Statistical Theory of Anomalous
  Fermion Number Nonconservation},''
  \href{http://dx.doi.org/10.1016/0550-3213(88)90133-2}{{\em Nucl. Phys. B}
  {\bfseries 308} (1988) 885--912}.

\bibitem{Khlebnikov:1996vj}
S.~Y. Khlebnikov and M.~E. Shaposhnikov, ``{Melting of the Higgs vacuum:
  Conserved numbers at high temperature},''
  \href{http://dx.doi.org/10.1016/0370-2693(96)01116-1}{{\em Phys. Lett. B}
  {\bfseries 387} (1996) 817--822},
  \href{http://arxiv.org/abs/hep-ph/9607386}{{\ttfamily arXiv:hep-ph/9607386}}.

\bibitem{Laine:1999wv}
M.~Laine and M.~E. Shaposhnikov, ``{A Remark on sphaleron erasure of baryon
  asymmetry},'' \href{http://dx.doi.org/10.1103/PhysRevD.61.117302}{{\em Phys.
  Rev. D} {\bfseries 61} (2000) 117302},
  \href{http://arxiv.org/abs/hep-ph/9911473}{{\ttfamily arXiv:hep-ph/9911473}}.

\bibitem{Rubakov:1996vz}
V.~A. Rubakov and M.~E. Shaposhnikov, ``{Electroweak baryon number
  nonconservation in the early universe and in high-energy collisions},''
  \href{http://dx.doi.org/10.1070/PU1996v039n05ABEH000145}{{\em Usp. Fiz. Nauk}
  {\bfseries 166} (1996) 493--537},
  \href{http://arxiv.org/abs/hep-ph/9603208}{{\ttfamily arXiv:hep-ph/9603208}}.

\bibitem{Burnier:2005hp}
Y.~Burnier, M.~Laine, and M.~Shaposhnikov, ``{Baryon and lepton number
  violation rates across the electroweak crossover},''
  \href{http://dx.doi.org/10.1088/1475-7516/2006/02/007}{{\em JCAP} {\bfseries
  02} (2006) 007}, \href{http://arxiv.org/abs/hep-ph/0511246}{{\ttfamily
  arXiv:hep-ph/0511246}}.

\bibitem{Morrissey:2012db}
D.~E. Morrissey and M.~J. Ramsey-Musolf, ``{Electroweak baryogenesis},''
  \href{http://dx.doi.org/10.1088/1367-2630/14/12/125003}{{\em New J. Phys.}
  {\bfseries 14} (2012) 125003},
  \href{http://arxiv.org/abs/1206.2942}{{\ttfamily arXiv:1206.2942 [hep-ph]}}.

\bibitem{Mukhanov:2007}
V.~F. Mukhanov and S.~Winitzki, {\em {Introduction to Quantum Fields in
  Classical Backgrounds}}.
\newblock Cambridge University Press, 2007.

\end{thebibliography}\endgroup

\end{document}